\documentclass[aps,prd,showpacs,amsmath,amssymb,reprint]{revtex4-1}
\usepackage{subfig}
\usepackage{graphicx}
\usepackage{float}
\usepackage{mathrsfs}
\usepackage{amscd}
\usepackage{natbib}

\begin{document}
\title{Network Gravity} 
\author{John Lombard}
\email[]{jml448@uw.edu}
\affiliation{University of Washington,\\ Seattle, Washington, 98195, USA}
\date{\today}

\begin{abstract}
We introduce the construction of a new framework for probing discrete emergent geometry and boundary-boundary observables based on a fundamentally a-dimensional underlying network structure. Using a gravitationally motivated action with Forman weighted combinatorial curvatures and simplicial volumes relying on a decomposition of an abstract simplicial complex into realized embeddings of proper skeletons, we demonstrate properties such as a minimal volume-scale cutoff, the necessity of a positive-definite cosmological constant-like term as a regulator for non-degenerate geometries, and naturally emergent simplicial structures from Metropolis network evolution simulations with no restrictions on attachment rules or regular building blocks. We see emergent properties which echo results from both the spinfoam formalism and causal dynamical triangulations in quantum gravity, and provide analytical and numerical results to support the analogy. We conclude with a summary of open questions and intent for future work in developing the program.
\end{abstract}


\maketitle

\section{Motivation}

At the quantum gravity scale, we do not expect spacetime to have a manifold structure \cite{Rovelli2015}. Geometry and associated operators such as volumes should be given by expectation values over a quantum superposition of states, which under a coarse macroscopic limit, return to a familiar metric construction. Frameworks such as string theory tend to work perturbatively around a background metric, and as a result may not provide insight into the foundational structure of spacetime in the regime where the background does not admit a metric topology \cite{Perez2013}. Canonical efforts such as the covariant spinfoam framework of loop quantum gravity attempt to address these questions more directly without presupposing a background. The spinfoam quantization of a constrained topological background field (BF) theory is based on an arbitrary simplicial decomposition of an underlying base manifold \cite{Baez2000}. The construction admits non-simplicial states--those where the quantum nature of the geometric operators only loosely impose the constraint conditions for proper geometries. These states are argued to peak to classical geometries in 6the appropriate limit; however, one must prescribe which dimensional BF theory one hopes to quantize, and as such, the literature is filled with work on 3d and 4d spinfoam models as quantizations of 3d and 4d gravity \cite{Perez2012}. Although these models are tremendous achievements toward a geometric and non-perturbative understanding of quantum gravity, here we take the perspective that absolute emergent dimensionality might be a property which a good theory of quantum gravity could hope to explain.

Causal dynamical triangulations (CDTs) are models which formalize the path integral notion of Hawking's `sum over geometries' approach to quantum gravity and rely on the layer-by-layer oriented construction of a triangulation under the evaluation of the Regge Action, the triangulation discretization of the Einstein-Hilbert action \cite{Ambjorn2001}. CDTs aim to shed light on such emergent dimensionality considerations. A long history of Euclidean triangulations is brimming with work detailing highly divergent path integrals or phases of emergent geometries with either Hausdorff dimension two or infinite \cite{Ambjorn2010}. CDTs are successful in finding a regime where a classical spacetime of spectral dimension four can be recovered \cite{Ambjorn2005}. It is the highly restricted nature of the paradigm with 4-simplex building blocks and constrained attachment rules which additionally forbid topology changes or branching geometries that ultimately admits this novel classical phase. Understanding why this limit arises or whether other conditions can generate similar behavior is the aim of current work in the field. 

The goal of the construction to follow is to investigate the limit of Euclidean classical geometries emerging from a fundamentally combinatorial network framework which does not presume the properties of an underlying triangulation. The field of emergent networks is a highly active area of research where the physical applicability of a model is often determined \emph{a posteriori} to the growth paradigm. Here, we attempt to make rigorous a stochastic growth paradigm which is designed specifically to probe questions in emergent simplicial geometry without guiding the growth structure `by hand', starting with a basic combinatorial structure and asking in what limits can it be demonstrated to contain substructures which approximate more familiar geometric constructions. If quantum geometry admits such a description, an analytical handle on emergent near-simplicial manifolds with non-simplicial defects may be obtained which will facilitate a better understanding of the semiclassical limit for the very strange quantum structures we expect at that scale. Observational evidence and phenomenological bounds for a fundamentally discrete structure to our universe could be ascertained by studying precisely such defects \cite{Hossenfelder2013,Hossenfelder2013a}. 

\section{Brief Summary}

The framework we develop has two complementary pictures. On one side, we define a space of states with boundaries built from a highly constrained, purely combinatorial structure. Every such state admits a representation as an embedded abstract simplicial complex with geometric realization at the skeletal levels. On this space, we define a Euclidean action which is a heuristic combinatorial analogue of the Regge action. We study the properties of this system and can sample the action-weighted space of states through traditional Markov Chain Monte-Carlo (MCMC) sampling techniques.

On the other side of the framework, we consider a space of unconstrained states consisting of embedded undirected networks with boundaries. We consider a stochastic process on the space of emergent networks and seek to sample the distribution of optimized final states for a finite horizon directed growth procedure under the evaluation of a scalar cost function.

We show through the explicit construction of the surjective covering that the space of embedded emergent networks can be mapped onto our combinatorial state space of interest. Equating the cost function with the combinatorial action, the optimization procedure on the embedded graph states translates into an importance sampling on the combinatorial space, generating an ensemble which is peaked around the minima of the action and allows for discrete topological observables to be computed against the states near classical fixed-points. 

\section{State Spaces}

We first take the opportunity to rigorously define the space of states, both on the combinatorial side with $\Psi_m$ representing the gauge fixed space of physical states, and in the emergent network picture with $\tilde{G}_m$ as the corresponding embedded graph space. Once this is established, we will demonstrate the covering between the spaces with an explicit construction of the map. We can initially set aside any boundary considerations, defining a generic bulk state instead. The addition of boundaries will be shown to constrain the spaces, but will not impede the general construction.

\subsection{Combinatorial State Definition}

A state $\phi_{m}\in\Psi_{m}$ is a rooted directed simple (excludes single-node loops and multiple edges degenerate between the same nodes) graph $g(V,E)$, 
where $V$ is the vertex set and $E$ is the edge set: 
\begin{eqnarray}
\{v_{i}\} &|& \{ v_{i}\in V \,\forall\, i\in[|V|]\}\ \, ,\\
\{(v_{i},v_{j})\} &|& \{ (v_{i},v_{j})\in E\,\forall\, v_{i}\neq v_{j}\}\,.
\end{eqnarray}
Let us denote rooted dipaths of edge cardinality $d$ (directed paths starting at a root vertex and containing $d$ edges in the graph) as $P_d$.
The state has maximal dipaths $P_m$, and is organized into `levels' indexed by $0\leq d\leq m$. 
Each level consists of the vertex subsets defined by the union of the terminals $T$ of $P_d$, 
\begin{equation}
V|_d=\cup_{i}v_{i}\,|\,\{ v_{i}=P_{d}^{T}\,\forall\, P_{d}\}\,. 
\end{equation}
The set of vertices at level $d=0$, $V|_0$, constitute the roots
of the graph. Edges are oriented outwardly from the roots toward higher
level vertices, and are constrained to connect only vertices at level $d$ with those
at level $d-1$.

The state is parameterized by a data structure $K^{*}=\sqcup_{d=0}^{m}K_{d}^{*}$
which defines the connectivity data, and a weight assignment $\omega_{\alpha_{d}}\in\mathbb{R}^{+}$
to each vertex at level $d$, where here $\alpha$ is a vertex label
on the indexing set $\alpha\in[|K_{d}^{*}|]$. 
Each $\alpha_{d}$ is a unique subset $\{\alpha_{0},\dots,\alpha_{0}\}$ of $d+1$
root vertices, which themselves are defined implicitly as the singletons
\begin{equation}
\alpha_{0}=\{\alpha\}\,\forall\,\alpha\in[|K_{0}^{*}|]\,. 
\end{equation}
Uniqueness is defined such that $|\alpha_{d}\cap\beta_{d}|<|\alpha_{d}|\,\forall\,(\beta\neq\alpha)$.
The reason for this particular vertex labeling will become clear in Eq.~\ref{eq:embeddedClique} when begin to define the projection from the embedded picture onto this combinatorial description.

$K_{d}^{*}=\sqcup_{\alpha}\alpha_{d}$
gives the connectivity structure through induced dipaths from the
previous $K_{d-1}^{*}$ down to the roots specified by each $\alpha_{d}$.
If one were to remove the vertices at $d=1$ and instead let the roots
continuously connect through the structure prescribed by $K_{1}^{*}$, one
would see that all vertices at level $d$ can be be traced back to cliques (complete subgraphs wherein every vertex is connected to every other vertex in the subgraph) of order $d+1$ in the
joined root graph. Note that not every clique between the roots generates
a vertex in the full state, however.

\begin{figure}[H]
\centerline{\includegraphics[scale=.3]{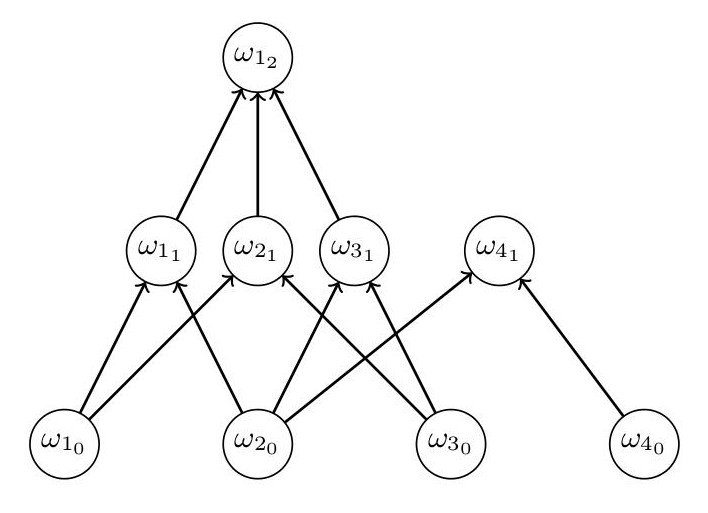}}
\caption{Representative of a Combinatorial State}\label{fig:Combo}
\end{figure}

In Fig.~\ref{fig:Combo}, we give an example of such a state $\phi_{m}$. To illustrate the $K^{*}$ structure of Fig.~\ref{fig:Combo}, we enumerate the components of the disjoint union at each level with a lexicographical ordering imposed on the roots as follows: 
\begin{eqnarray}
 K_{0}^{*} &=& \{(\{1\},1),(\{2\},2),(\{3\},3),(\{4\},4)\} \nonumber \,; \\
 K_{1}^{*} &=& \{(\{1,2\},1),(\{1,3\},2),(\{2,3\},3),(\{2,4\},4)\} \,;\\
  K_{2}^{*} &=& \{(\{1,2,3\},1)\} \nonumber \,.
\end{eqnarray}
In addition to the defining properties of this parameterization, there are a host of combinatorial inequalities that must be enforced to restrict to the state space
we are interested in. For a discussion on such inequalities, see App.~\ref{sec:ComboIneq}.

In short, the combinatorial structure must be such that for every state $\phi_{m}$ there exists at least $1$ abstract simplicial complex embedded in $\mathbb{R}^{m}$ whose $d$-skeleton is given by $K^{*}_d$, and admits embedded simplicial volumes with magnitudes given by the corresponding $\omega_{\alpha_{d}}$ weights. This will be illustrated in more detail by construction with respect to the embedded paradigm below. 

\subsection{An Embedded State}
Having defined a combinatorial state, we now introduce a complementary embedded graph state which will be mapped onto our combinatorial space.

Let $g(V,E)$ be an simple undirected graph.

Let $\chi_{m}\,:\, g\mapsto\tilde{g}_{m}$ be a fixed embedding of $g$ into
$\mathbb{R}^{m}$ such that $\chi_{m}(V(g))$ is injective and $\chi_{m}(E(g))$
is the collection of unique geodesic paths between the connected nodes
with path lengths given by the $L_{2}$ norm between the nodes from the induced metric
on the ambient space (i.e. straight line graph embeddings with no
degenerate points). 

\begin{figure}[H]
\centerline{\includegraphics[scale=.25]{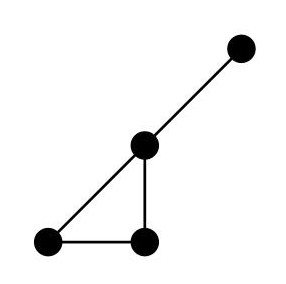}}
\caption{Representative Embedded Graph State Corresponding to Fig.~\ref{fig:Combo}\label{fig:Embd}}
\end{figure}

We consider $\tilde{g}_m =\tilde{g}_m (g,\chi_m)$ to be an embedded graph state. 

\subsection{From Embeddings to Combinatorics\label{sec:duality}}
From $\tilde{g}_{m}$, we will explicitly construct a map $\mu$ which projects onto the combinatorial space.

\begin{figure}[H]
\centering
\subfloat{(a)\includegraphics[scale=4]{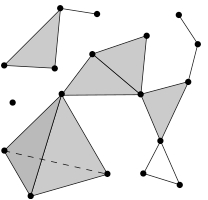}}(b)\subfloat{\includegraphics[scale=3]{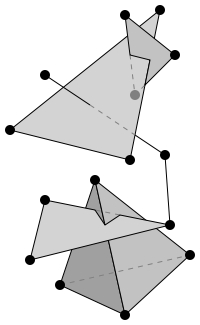}}
\caption{A Simplicial Complex (a) and a Clique Complex (b), both Embedded in $\mathbb{R}^{3}$}\label{fig:Abstract-and-'Proper'}
\centering
\end{figure}

First, given $\tilde{g}_{m}$, we construct its ordered clique complex. Let
a complete subgraph of order $d+1$ be denoted $\alpha_{d}=\{v_{i}\cdots v_{j}\}$,
where $\alpha$ is a labeling on the set of complete subgraphs at
fixed order. Every $\alpha_{d}$ is in bijection with a combinatorial simplex of order $d$. We form the ordered clique complex by taking the disjoint union of simplexes associated with each complete subgraph. That is, define the clique skeleton at order $d$ to be 
\begin{equation}
K_{d}=\sqcup_{\alpha}\alpha_{d}\,|\,\{\alpha\in[|K_{d}|] \}\,. 
\end{equation}
This complex is an embedded abstract simplicial complex.
A traditional simplicial complex has stringent requirements on the intersections of its simplicies such that the intersection of two simplicies occurs as a subset of the union of their boundaries which is also a lower dimensional simplex. An abstract complex has no such restrictions. Additionally, not every complete subgraph of the edge skeleton in a simplicial complex actually forms a simplex, as opposed to a clique complex. Fig.~\ref{fig:Abstract-and-'Proper'} illustrates this difference.

We introduce a map 
\begin{equation}\label{eq:OmegaMap}
\Omega\,:\,\tilde{\alpha}_{d}\rightarrow\mathbb{R}^{+} 
\end{equation}

which computes the simplicial $d$-volume of the embedded cliques $\tilde{\alpha}_{d}$.
As discussed, the cliques may not have a well defined geometric realization in the complex under $\chi_{m}$ and as a result we are unable to define a proper volume for them.
To rectify this, we generate a set of individually embedded cliques by restriction, compute the volumes where there is guaranteed to be such a proper notion, and map the information back to the full state through coherent indexing.

Let $R_{\alpha_{d}}$ be a restriction map such that 
\begin{equation}
R_{\alpha_{d}}(g(V,E))=g(\alpha_{d},E|_{\alpha_{d}})\,.
\end{equation}

Each individually embedded clique is simply 
\begin{eqnarray}
 \chi_{m}(R_{\alpha_{d}}(g))&\equiv&\tilde{\alpha}_{d} = (\overbrace{p_i \cdots p_j}^{d+1})\, \nonumber \\
 p_i&=&\chi_m(v_i) \in \mathbb{R}^m \, , \label{eq:embeddedClique}
\end{eqnarray}
using the same indexing on $\alpha_d$ to carry the labeling data between the maps.

Let $\omega_{\alpha_{d}}\equiv\Omega(\tilde{\alpha}_{d})$. 

Define the following structure as a skeletal subset: 
\begin{equation}\label{eq:K'}	
K'_{d}\equiv\ \sqcup_{\alpha}\alpha_{d}\, | \,\{ \,\omega_{\alpha_{d}}>0\} .  
\end{equation}
This produces $d$-skeletons which have elements that, when considered in isolation, have proper $d$-volume. That is, the embedded subgraphs do not span a lower-dimensional hyperplane, or contain too many vertices to be linearly independent given the dimensionality $m$ of the ambient space.

Now define a new map $I$ based on intersection pruning: 
\begin{eqnarray}\label{eq:intersection}
I\,:\, K'{}_{d}&\mapsto& K_{d}^{*}\\ 
K_{d}^{*} & = & \sqcup_{\alpha} \alpha_{d} \,| \, \{ \Omega_{d}(P_{\tilde{\alpha}_{d},\tilde{\beta}_{d}})=0 \,\forall \,\beta_d\neq\alpha_d\in K'_{d}   \} \nonumber \,,
\end{eqnarray}
where $\Omega_{d}$ is the $d$-volume associated with the convex
interior of the intersection polytope $P_{\tilde{\alpha}_{d},\tilde{\beta}_{d}}$ defined by 
\begin{equation}
 P_{\tilde{\alpha}_{d},\tilde{\beta}_{d}}  =  conv(\tilde{\alpha}_d)\cap conv(\tilde{\beta}_d) \, .
\end{equation}
This volume can be computed up to arbitrary precision by considering an $\epsilon$-fine $d$-triangulation $\Delta^{\epsilon}_d$ of $P_{\tilde{\alpha}_{d},\tilde{\beta}_{d}}$,
and acting with $\Omega$ distributively on the simplexes in the triangulation, 
\begin{equation}
\Omega_{d} \equiv \sum_{\delta\in\Delta^{\epsilon}_d(P_{\tilde{\alpha}_{d},\tilde{\beta}_{d}})}\Omega(\delta) \,. 
\end{equation}

By convexity, the two $d$-simplexes being tested for intersection can clearly not have an intersection polytope that spans a higher dimensional ambient space. However, if the intersection polytope spans a lower dimensional ambient space with dimension $d'<d$, then $\Omega_{d}$ is naturally $0$.  

In practice, $I$ is a binary intersection test distributed over all
cliques in a skeleton, and does not use triangulation volumes but
other faster and more robust convex intersection algorithms instead, such as solving a linear-programming problem on the convex hull of the Minkowki difference of the simplexes.

Given our suggestive notation, it should be apparent that each $K_{d}^{*}$, which we term a `proper pruned skeleton', along
with the associated weights $\omega_{\alpha_{d}}\,\forall\,\alpha_{d}\in K_{d}^{*}$,
make up precisely the combinatorial data for the state $\phi_{m}$ which satisfies our set of constraints by construction,
and can be used to form the vertex-labeled digraph structure in the un-embedded combinatorial description. 

We take the above sequence of maps to define the projection
\begin{eqnarray}
\mu\,:\,\tilde{g}_m \mapsto\phi_m \, . 
\end{eqnarray}

\subsection{Embedding the Combinatorial Data}
Just as $\mu(\tilde{g}_m)\mapsto \phi_m$, we can implicitly define a representative minimal embedding back from the combinatorial data into $\mathbb{R}^{m}$.

Out of the infinite family of embeddings which satisfies this property, let $\eta_m$ be one representative fixed embedding in $\mathbb{R}^m$ such that 
\begin{equation}\label{eq:eta}
 \phi_m = \mu\circ\eta_m\circ\mu(\tilde{g}_{m})\,.
\end{equation}

We note that due to the non-injectivity of $\mu$, 
\begin{equation}
\eta(\phi_m) \neq \tilde{g}_{m}
\end{equation}
 in general. Even after removing the vertex/root labels on both sides of the map, there are multiple states with different embedded intersection properties that can yield the same combinatorial data defining $\phi_m$. However, for every pairing $(g,\chi_{m})$ there exists a unique combinatorial state up to inherited gauge equivalence.

For a discussion of some of the gauge symmetries of this theory, see Sec.~\ref{sec:Symm}. An example of the equivalence classes under gauge that we are referring to would be those graphs equivalent up to isometry in the ambient embedding space, $ISO(m)$, or up to vertex automorphism in their graph description, $Aut(g)$.

\subsection{Physical Covering Space}

We have constructed a map $\mu$ such that a combinatorial state $\phi_m$ can be parameterized by entirely by $(g,\chi_{m})$. 

A state $\tilde{g}_m$ lives in the unrestricted space of all possible embeddings of all possible graphs (of the type of graph/embedding pairing we consider). Denote this space $\tilde{G}'_m$. This space is dense in combinatorial states $\phi_m$ by construction under $\mu$ as a multi-cover into a combinatorial space we denote $\Psi'_m$. However, we are interested in the unique gauge-fixed equivalence classes of combinatorial states, the physical space $\Psi_m$. 

Take the following quotient map as a projection onto the combinatorial gauge-fixed base space, identifying the multi-cover into unique equivalence classes.

\begin{equation}
 Q\,:\,\Psi'_{m}/{\sim} \rightarrow \Psi_m
\end{equation}

We can formally define a pull-back of $Q$ to the embedded space by using $\eta_m$ such that the following diagram commutes: 

\begin{equation}
 \begin{CD}
   \tilde{G}'_m @> \mu >> \Psi'_m \\
 @V\exists V \tilde{Q} V @VV Q V \\
 \tilde{G}_m @> \eta^{-1}_m >> \Psi_m
 \end{CD}
\end{equation}

In the lower map, $\eta^{-1}_m$ provides an isomorphism between the reduced covering space and the physical combinatorial state space. 

We note that no such gauge fixing is known to exist at the time of writing. As is often the case, we trade difficult to characterize rigid constraints in the combinatorial description for difficult to characterize inherited symmetries in the embedded description. Nevertheless, it is computationally easier to handle the redundancy in the multi-cover than it is to probe the constrained space directly.

\subsection{Formal Path Integrals}
In Sec.~\ref{sec:Action} we explicitly define the action $S(\phi_{m})$ on the combinatorial space $\Psi_m$.
Our action is decomposable into the combinatorial levels, and as a result, we can formally define a Euclidean path integral for our partition function as follows: 

\begin{equation} 
Z_{m} = \int  \mathscr{D}\phi_{m} \,\exp{\sum_{d=0}^{m}-S(K^{*}_{d},\omega_{\alpha_{d}})}
\end{equation}

Regarding numerical work, sampling the full space directly is difficult due to the constraints. Nevertheless, with the projection $\mu$, we can now sample a proxy-space instead and be guaranteed to be sampling the entirely of our space of interest with the same action: 
\begin{equation}
\tilde{Z_{m}}=\int  \mathscr{D}\tilde{g}_m \,\exp{(-S(\mu(g,\chi_{m})))}
\end{equation}

We are not guaranteed, however, that a uniform sampling of $\tilde{G}_{m}'$ will have its distribution preserved under the map. Without further study, we cannot say that in the limit of a sufficiently well mixed Markov chain, the measure we intend to sample through a Metropolis filter on the embedding space is truly the measure we receive after we map to the combinatorial space, prohibiting a full quantum simulation of the combinatorial space of interest. An outstanding computation which would enable such a full sampling would be to demonstrate the convergence of the measures under $\mu$ such that
\begin{equation}
 P[d_{TV}(\mathscr{D}\phi_{m},\mu(\mathscr{D}\tilde{g}_{m})) > \epsilon ] < 1 - \delta \, ,
\end{equation}
for small parameters $(\epsilon,\delta)$, uniform measures on the base spaces, and total variation distance between the measures $d_{TV}$. Such a proof is the object of current work.

Seeking the action minima, however, is entirely within our capabilities as they are trivially the same on both sides of the map, and a strongly driven optimization problem on the embedded side yields a combinatorial distribution highly peaked around classical states of interest. The majority of this discussion will work in such a paradigm. 

\subsection{Boundaries}

There exist special subsets of the states which are immutable
in the state sum. These are deemed to be the state boundary, $\partial \tilde{g}_{m}$, wherein the embedded graph subset is unable to be altered in either its structure or embedding data, relative to the other vertexes in the boundary. The combinatorial equivalent is given in terms of a minimal fixed $K^{*}$ structure that cannot be altered (and induces a set of constraints on the weights at the nodes of that structure), plus an additional set of finite constraints on the weights associated with $K^{*}_1$. See App.~\ref{sec:ComboIneq}. 

\begin{figure}[H]
\centerline{\includegraphics[scale=4]{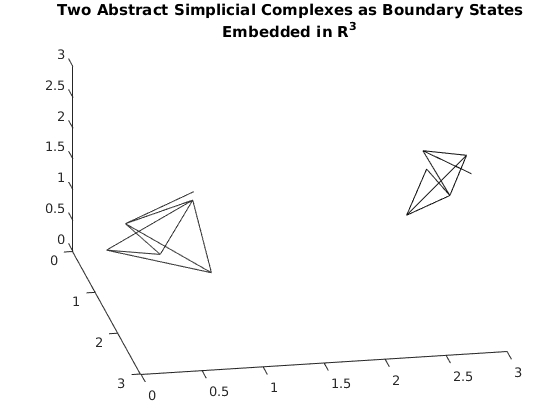}}
\caption{Example of an Initialized Boundary State}
\label{fig:Boundary}
\end{figure}

In the embedded picture, a state boundary is characterized by its number of convex-hull disjoint path-components. That is, we can have a single closed boundary made of one path component, a $2$-boundary system where the state sum is over bulk geometries between initial and final configurations, or a multi-boundary state given by $n$ convex-hull disjoint components embedded in the same ambient space. Each component is assumed to have the same geometric realization structure encoded in the combinatorial equivalent, such that given a boundary component $B(g,\chi_{m})\in\partial \tilde{g}_m$,
\begin{equation}\label{eq:boundary}
 B(g,\chi_{m}))\sim\eta_m\circ\mu(B(g,\chi_m))) \, ,
\end{equation}
 where $\sim$ denotes equality under gauge equivalence. The initial embedding of the boundary data establishes the relative orientation of the substructures in the boundaries and any intrinsic length scales of the system, which are also represented in the set of combinatorial constraints. 
An example of a two path component boundary system initialization can be seen in Fig.~\ref{fig:Boundary}.
In addition to the protected nature of these
boundaries under the Markov process, the intersection map $I$ is also
boundary preferential and is modified as follows:
\[
 K_{d}^{*} = \sqcup_{\alpha} \alpha_{d} \left\{
\begin{array}{ll}
      \forall \alpha_{d}\in\partial\tilde{g}_{m} \tag{\ref{eq:intersection}\'} \\
      \Omega_{d}(P_{\tilde{\alpha}_{d},\tilde{\beta}_{d}})=0 \,\forall \,(\beta_d\neq\alpha_d)\in K'_{d} \,.  
\end{array} 
\right.
\]
\section{Ambient Space}

In order to probe emergent geometry, our discrete model must make contact with discretizations of manifolds. Although our model can be formally generalized to work with cellular decompositions of manifolds, we gain a robust computational toolkit with a restriction to convex/simplicial structures. With the simplicial approximation theorem in mind, we choose to work in a flat space and impose the geodesic embeddings of our graph edges in order to restrict to abstract simplicial complexes instead of abstract cellular complexes, without loss of generality for manifold approximation \cite{Hatcher2002}. 

We avoid pathological flat ambient spaces, for if they cannot admit topologically `reasonable' immersions of manifolds, then we have no hope to grow approximations to those structures. Lastly, we would like the ability to unambiguously define unique edge lengths without the use of an additional choice for periodically identified spaces, and we would like the full space available for an embedding as to avoid spaces with any singularities which could localize the graphs, establish extrinsic length scales, or otherwise impact the state geometry. As a result, we work in $\mathbb{R}^{m}$. 

In the emergent network picture, we aim to explore the impact that the global embedding space has on stochastically grown networks between provided boundary states. From the combinatorial side, this is a restriction on the maximum number of levels $m$  available in the combinatorial tree. Although we work at fixed embedding dimension, nothing in our construction is explicitly dependent on the ambient dimension--it is simply a constraint that can be taken to be infinite, or much larger than any intrinsic dimensionality of the boundary states. This control is important in the context of quantum gravity, as we would hope to allow for bulk states which may explore arbitrary dimensional configurations in the state sum. From a holography perspective, we can then ask questions like: 
\begin{quote}
Given a boundary state which is a triangulation of $S^2$ embedded in $\mathbb{R}^3$, is the optimal bulk state a triangulation of $B^3$ in the interior? If we embed in $4$ or higher dimensions, does the high dimensional bulk data still lie in the vicinity of a much lower dimensional sub-triangulation?
\end{quote}
The later opens up questions that breach the realm of probing the manifold learning hypothesis  \cite{Fefferman2016}.
For computational purposes and to explore the effects of a finite embedding space on network growth with fixed boundary, we begin each investigation with the minimal embedding dimension that accommodates the boundary data in accordance with Eq.~\ref{eq:boundary} and explore for asymptotic behavior in the large ambient limit.

\section{Markov Process and Metropolis Algorithm\label{sec:Markov}}

Within the paradigm of optimization of the embedded networks, we now describe the finite horizon Markov process by which the network undergoes an evolutionary step $t\in \mathbb{N}$.

\begin{figure}[H]
\begin{center}
\includegraphics[scale=4]{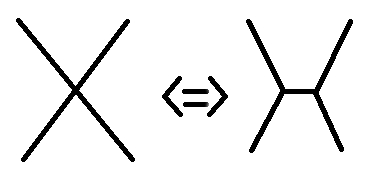}\caption{An Example Markov Move and its Inverse}\label{fig:nodalSplit} 
\end{center}
\end{figure}

Let $M_{t}$ be a set of available moves $m_{i}\in M_{t}$ indexed by $i$ at step $t$ in the evolution. The set of available moves have been chosen such that every move has an inverse which admits detailed balance, and is heuristically chosen such that the algorithm can freely sample across a large space of admissible states. We establish a minimal set of perturbations of $g$ and $\chi_m$ which cover $\tilde{G}'_m$:
\begin{enumerate}
 \item Nodal addition and subtraction
 \item Edge addition and subtraction
 \item Nodal perturbation in the ambient space 
\end{enumerate}
Of course, the moves are restricted to respect the static boundaries and simplicity conditions imposed in by our graph definition, as well as vertex injectivity.
Compositions of these moves enable access to all boundary-respecting states in the space. We also include global moves to promote ergodicity and rapid mixing which alter both $g$ and $\chi_m$ simultaneously, such as nodal splitting/recombining, edge splitting/recombining, and multiple simultaneous node additions/subtractions and edge additions/subtractions. See Fig.~\ref{fig:nodalSplit} for an example of such a nodal splitting/recombining move. The joined/split nodes are uniformly randomly relocated in a compact subset of the ambient space.

We take the Markov process to be a sequence $(g_{1},\chi_{m}^{1})\dots(g_{n},\chi_{m}^{n})$ of states where the maps from $(g_{i-1},\chi_{m}^{i-1})\mapsto(g_{i},\chi_{m}^{i})$ are the identity in the subspace untouched by the perturbation. Although we could select embeddings other than the identity which still preserve the full combinatorial data in the complementary subspace, we choose the identity for convenience. Picking such a particular embedding chain is fixing some of the gauge inherited from the ambient freedom. 

Let $p_{t}^{i}$ be the probability for performing move $m_{i}$:
\begin{equation}
p_{t}^{i}=\frac{1}{|M_{t}|}\,\forall\, i\,.
\end{equation}

Given a boundary initialization, an annealing temperature $T_{t}\in\mathbb{R^{+}}$ and an annealing constant $\beta=(0,1]$ are established at the start of an algorithm of annealed descent. As the simulation termination condition is the completion of a finite number of steps, we set the annealing to be sensitive to the simulation horizon, $t_{max}\in\mathbb{N}$.

$\beta$ controls how much of the simulation is in a deterministic decent regime with zero simulation temperature. We desire that large initial fluctuations allow for perturbations into a boundary-biased `random' configuration, wherein thereafter the simulation begins to settle on a local branch minimum before it strictly descends to the optimal configuration of that branch. In this way we achieve the freeze-out sampling of minima of the state space. Sampling through repeated runs of the algorithm builds a space of semiclassically stable local-minima which, after being pruned for accidental degeneracies or repeated sampling by selecting only a single representative from unique equivalences classes, can be re-weighted by the action and used in a traditional formalism for partition function observables. 

The annealing temperature is lowered geometrically as follows:
\begin{equation}
T_{t+1}=T_{t}\,(1-\frac{t}{\beta\, t_{max}})\,(1-\theta(t-\beta\, t_{max}))\,,
\end{equation}

with $\theta$ being the standard Heaviside function. The state then enters the sampling algorithm. 
\begin{description}
\item [{Algorithm}] The action $S_{t}$ is computed according to Eq.~\ref{eq:Action}. $M_{t}$ is determined. A move $m_i$ is uniformly randomly selected and performed, while incrementing the step counter $t\mapsto t+1.$ The action $S_{t+1}$ is then computed. If $S_{t+1}<S_{t}$, the move is accepted, relevant data is recorded and simulation parameters are updated, and the algorithm repeats. Otherwise, moves which increase the action are accepted with conditional probability $\frac{\pi_{t+1}}{\pi_{t}}$ of the form 
\begin{equation}
 \frac{\pi_{t+1}}{\pi_{t}}=\exp{\frac{-(S_{t+1}-S_{t})}{T_{t}}}\,.
\end{equation}

Rejected moves return the state to its previous configuration. The algorithm continues to run until an exit condition is met.
\end{description}

\section{Combinatorial Gravity}
Our goal is to understand the roll of gravity in emergent geometry from a combinatorial perspective. However, our traditional notions of gravity are not well defined for a combinatorial framework. The Einstein-Hilbert action on a manifold $M$ is defined as follows: 
\begin{equation}
S_{EH} = \int_{M} \sqrt{-g} (R - \Lambda)  \,,
\end{equation}
where the Jacobian is built from the metric determinant $g$, $R$ is the Riemannian Ricci scalar, and $\Lambda\in\mathbb{R}$ is the cosmological constant. The volume-weighted curvature form of this action can be effectively discretized for triangulations of a $d$-dimensional manifold, giving us the action of Regge calculus:
\begin{equation}
 S_{R}=\sum_{h}V_{h}\epsilon_{h}+\Lambda\sum_{\sigma}V_{\sigma}\, ,
\end{equation}
where $d$-simplexes are indexed by $\sigma$, $(d-2)$-simplexes (hinges where curvature is concentrated) are indexed by $h$, simplicial volumes $V$ replace the continuum Jacobian, and deficit angles $\epsilon$ replace the Ricci curvature  \cite{Friedberg1984}. 

We would like to understand what are the minimal prescriptions necessary to generate geometric structures. To make contact with the nearest discrete gravitational formalism of Regge Calculus, we need to utilize combinatorial notions of volumes and curvatures which we can package in an effective combinatorial action. We do not have the rigid structures of even simplicial complexes at our disposal, and as a result, we seek to carefully construct a `network gravity' formalism which admits analogous structures in the appropriate limit, while maintaining agnosticism with respect to fundamental building blocks and attachment rules. 

\section{Proper Pruned Skeletons\label{sec:Pruning}}

In Sec.~\ref{sec:duality} we outlined explicitly the map from the embedded space to the combinatorial space, including the pruning and excision process which defined for us the $K^{*}$ structure. From the embedded perspective alone, the states generically admit a large amount of geometric defects. This is a desired feature, for if the model aims to probe emergent discrete geometry at all scales, there should neither be constraints on the dimension of the building blocks nor their matings. Macroscopic consistency would only demand that in a regime where we expect general relativity to be applicable, the emergent description should be one which may approach a realized triangulation of an underlying manifold.

\begin{figure}[H]
\centerline{\includegraphics[scale = .3]{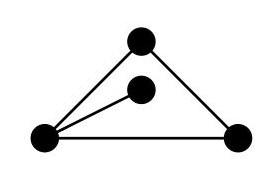}}
\caption{Admissible 1D Geometric Defect on a 2-Simplex Embedded in $\mathbb{R}^2$\label{fig:Defect}}
\end{figure}

As we do not have a dimensional specification, we must allow for a description of a state that measures its near-triangulation structure at each simplicial level. Breaking down a generic embedded graph state into the $K^{*}_{d}$ skeletons is precisely such a structure. Each skeleton alone is an embedded simplicial complex with geometric realization of all of its simplicial elements that can be used as a measure against a full triangulation, with the defects in the global state carried by the decomposition structure. The state is globally a superposition of abstract simplicial skeletons that each admit volumetric embeddings, where the non-geometric data at each skeletal level only comes from lower-dimensional structures. For example as shown in Fig.~\ref{fig:Defect}, we may admit an embedding of a $2$-simplex that has a $1$-simplex in its interior, giving a $1$-dimensional defect on the embedded $2$-skeleton, but by construction there can be no non-geometric data of dimension $2$ or higher paired with our $2$-simplex.

Generic realizations which take abstract simplicial complexes to simplicial complexes do not have the embeddability and volume constraints that we impose \cite{Hatcher2002}. We do not have the freedom to generically untangle the combinatorial network to force a geometric embedding of every element, or to find an embedding space of high enough dimension to admit a full realization, and instead turn to our in-place decomposition which admits a consistent graph structure for which we show that we can ascribe proper volumes and curvatures--essential tools for probing an effective gravitational action.

\section{Weightings\label{sec:Weightings}}

As mentioned, the weightings $\omega_{\alpha_{d}}$ assigned to a state's complete subgraphs are based on the embedded simplicial volumes of those subgraphs. Our $K^{*}$ structure guarantees that we can properly define a $d$-volume for each $d+1$-clique in bijection with a $d$-simplex. 

\subsection{Proper d-Skeleton Weights} 

$K^{*}_{1}$ corresponds to the $\alpha$-indexed list of 1-simplex edges in the embedded graph state. Each edge $\alpha_{1}$ is then assigned a weight $\omega_{\alpha_{1}}$ given by its $L_{2}$ coordinate length along the graph from the inherited metric of the ambient space.

This construction continues for all dimensional simplexes, with $K^{*}_{2}$ associated with the disjoint union of $2$-simplexes and weights $\omega_{\alpha_{2}}$ given by their corresponding triangle areas, etc.

In general, the simplicial $d$-volume is given by the Cayley-Menger determinant, with the $L_{2}$ norm used to compute the edge lengths between the embedded nodes for each edge in the simplex \cite{Sommerville1958}. We produce $B_{\alpha_{d}}$, a symmetric off-diagonal matrix of squared edge lengths, by indexing over all embedded vertexes $p_{i}\in \tilde{\alpha}_d$.

For $d>0$: 
\begin{subequations}
\begin{eqnarray}
\Omega(\tilde{\alpha}_d)&=&\sqrt{\frac{(-1)^{d+1}}{2^{d}d!^{2}}\det(\hat{B}_{\tilde{\alpha}_d})} \,; \label{eq:omega} \\
\hat{B}_{\tilde{\alpha}_d}&=&
\begin{bmatrix}
0 & 1 &\cdots& 1\\
1\\
\vdots &  & \mbox{\large\ensuremath{B_{\tilde{\alpha}_d}}}\\
1
\end{bmatrix}
, B_{\tilde{\alpha}_d}=\|p_{i}-p_{j}\|_{2}^{2}\,.
\end{eqnarray}
\end{subequations}

Note that for $d>m$ or subgraphs spanning hyperplanes of dimension less than $d$, all $\omega_{\alpha_{d}}$ will necessarily be zero as the coordinates cannot span the necessary vector space. We note that although the weights are derived from simplicial volumes, they are defined to be unitless. We simply take the simplicial volume as a combinatorial magnitude which respects the image of the volume map into $\mathbb{R}^{+}$.

\subsection{0-Skeleton Weights}

As the vertex set does not have any simplicial volume, we assign a single uniform weighting $\omega_{\alpha_{0}}\equiv \omega_0 \in\mathbb{R}^{+}$ to each node, as they are geometrically equivalent point objects. This marks the first free parameter of the theory, and its consequences will be discussed at length. 

\subsection{Curvatures}

As we do not have a triangulation structure, we have no way of pre-associating curvature to specific substructures, such as those found in the Regge case when employing deficit angles \cite{Miller1997}. In this network formalism, we proceed with the intent that $d$ itself is a true emergent observable and utilize a combinatorial curvature which generalizes to give a measure of discrete curvature over all simplicial scales. See App.~\ref{sec:Knill-Curvature} for a description of Knill Curvature, a purely combinatorial curvature measure often used in network theory that we justify is too topological for our purposes as it fails to take into consideration the Euclidean volumes of the emergent realized geometry.

\subsubsection{Forman Curvature}

A combinatorial curvature associated with any cell of a quasi-convex cellular complex can be described by the dimensionless Forman scalar curvature derived from the Bochner-Weitzenbock decomposition of the combinatorial Laplace operator \cite{Form2003}. A regular CW complex is quasi-convex if for every pair of $d$-cells $(\alpha,\alpha')$ and $d-1$ cell $\gamma$, 
\begin{equation}
\gamma\subset(\bar{\alpha}\cap\bar{\alpha}')\Rightarrow(\bar{\alpha}\cap\bar{\alpha}')=\bar{\gamma}\,. 
\end{equation}
We note that a simplicial complex is a subset of this larger class, and that this scalar curvature can be applied to give a curvature measure on each $\alpha_{d}\in K^{*}_d$, as $K^{*}_d$ satisfies quasi-convexity by construction with the map $I$ in Eq.~\ref{eq:intersection}. 

This definition of curvature is a weighted combinatorial curvature which depends on the near-nonlocal data of the $\alpha_{d\pm1}$ neighbors ($\alpha_{d}\subset\alpha_{d+1}$ or $\alpha_{d-1}\subset\alpha_{d}$) of the original simplex, and all of the associated weights. Formally, the list of weights is an arbitrary assignment into $\mathbb{R}^{+}$ inherited from an inner product on the cellular chain complex and can be provided by topological data, selected from some distribution, or set as a standard set of weights by taking them all to be unity. Here, we assert by model ansatz that the weights be precisely the embedded weights provided by the $\Omega$ map in Eqn.~\ref{eq:omega}, which allows this measure of curvature to account for the relative sizes of the simplicial pieces, along with their connection data. Such an assignment was suggested by Forman as an approach to possibly connect the study of curvature on embedded combinatorial manifolds as cellular decompositions to their Riemannian analogues, imbuing the combinatorial structure with a sense of the intrinsic geometry of the cellular pieces \cite{Form2003}. 

For $d=1$, the Forman curvature is strongly analogous in homological properties to the Riemannian Ricci curvature when the edges are considered as part of a cellular decomposition of a manifold, making them a good analog for gravitational curvature in this formalism. The higher curvature terms correspond to a generalization of the scalar curvature for the higher dimensional cells.

The Forman curvature is given as follows: 
\begin{widetext}
\begin{equation}
F(\alpha_{d})  =  \omega_{\alpha_{d}}\left\{ \sum_{\alpha_{d+1}\supset\alpha_{d}}\frac{\omega_{\alpha_{d}}}{\omega_{\alpha_{d+1}}}+\sum_{\alpha_{d-1}\subset\alpha_{d}}\frac{\omega_{\alpha_{d-1}}}{\omega_{\alpha_{d}}}\right.
    - \left.\sum_{\tilde{\alpha_{d}}\neq\alpha_{d}}\left|\sum_{\substack{\alpha_{d+1}\supset\alpha_{d}\\
\alpha_{d+1}\supset\tilde{\alpha_{d}} 
} 
}\frac{\sqrt{\omega_{\alpha_{d}}\omega_{\tilde{\alpha_{d}}}}}{\omega_{\alpha_{d+1}}}-\sum_{\substack{\alpha_{d-1}\subset\alpha_{d}\\
\alpha_{d-1}\subset\tilde{\alpha_{d}}
}
}\frac{\omega_{\alpha_{d-1}}}{\sqrt{\omega_{\alpha_{d}}\omega_{\tilde{\alpha_{d}}}}}\right|\right\} \,.
\end{equation}
\end{widetext}
We note that for a uniform point weight, $F(\alpha_{0})=0$. We also note that unlike the Riemannian case where there is no intrinsic curvature, Forman curvature can be ascribed to edges of a $1$-dimensional triangulation. This is a known obstruction to having this curvature measure be more closely aligned with a smooth equivalent \cite{Form2003}. In our case, we can recover the right intrinsic curvature behavior in $1$-D when the point volume is zero, however. This has led us to hypothesize that once the adjacency structure is set at a regulated value of $\omega_0$, there may be a renormalization of the point volume back to the physical value that can be accomplished by an secondary annealing, holding $K^{*}$ fixed. Investigations are forthcoming on this front.

\section{Action \label{sec:Action}}

The action is an effective model based on the $K_{d}^{*}$ slicing of the network. For each slicing, a Regge-like action is implemented, with the volume form given by the combinatorial map of the embedded simplicial weights, and the curvature form given by the Forman curvature. The action includes a sum over all possible proper pruned $d$-skeletons, and for a finite network, itself necessarily contains a finite number of terms. 
\begin{equation}
S(\phi_m)=\sum_{d=0}\xi_{d}\sum_{\alpha_{d}\subset K_{d}^{*}}\omega_{\alpha_{d}}(F(\alpha_{d})+\Lambda)\,.\label{eq:Action}
\end{equation}

Here, the  $\xi_{d}$ are coupling constants which differentially weight the slices of the network. This can be repackaged as a tower of coupling constants for each higher curvature term, similar to those found in a Lovelock theory of gravity where a sum over all Euler densities in $d$-dimensions constitutes the most general gravitational action \cite{Padmanabhan2013}. A near-analog to the Gauss-Bonnet theorem for the Forman curvature has been established for a combinatorial choice of weights, which indicates that the curvature has similar topological properties to the smooth Gaussian curvature for certain cellular dimensions \cite{Form1998}. To date, no such analog has yet been generalized for the embedded weights we have utilized here however. There is no modeling constraint at this time which guides our couplings, although further investigation may lead to restrictions in order to generate particular behaviors of interest. Uniform weighting over all $d$-skeletons occurs with $\xi_{d}=1\,\forall\, d$, and is the only 
distribution we have found experimentally thus far which demonstrates the phenomenology we are interested in. 

The cosmological constant term $\Lambda$ is a dimensionless scalar offset to the Forman curvature representing a uniform background combinatorial curvature present over all skeletons. We will show at length the effect this term has on classical states.

\section{Regulation}

The action is intrinsically not positive-definite. As such, in the descent paradigm, the global minimum may be unbounded with network growth. Two of the free parameters we have introduced so far, the point volume and the cosmological term, act as regulators against such configurations.

\subsection{Point Volume and Minimum Combinatorial Weight}

As a simplicial volume, a natural choice from a geometric perspective may be to take $\omega_0=0$. However, the Forman curvature dictates that the weights must be strictly positive. If we were to force the situation of a zero point volume, we additionally find that there is an equivalence class of actions under the addition of disconnected points, where $\Delta S=0$. This implies an identification between the absolute empty state and a state of infinite point density under the action.

While one solution to break this degeneracy would be to simply remove any disconnected points as `non-participating' elements of the state, the geometric significance of a state which is infinitely dense with disconnected points is lost. Underneath the cover of points, a non-trivial network may have non-zero action due to the $K_{d}^{*}$ slicing of the network. But in practice, there is no way to discern whether two points are connected in the infinitely dense sea, indicating that such an equivalence class is disfavored geometrically.

\subsubsection{Pruning Modification \label{sub:Pruning-Modification}}

The presence of a positive definite point volume sets a minimum combinatorial length scale to the system. We assert that a combinatorial volume with $\omega_{\alpha_{d}}\leq\omega_0$ should not be able to be resolved, and require that our pruning procedure for determining geometric realization respect the point-volume as a lower combinatorial-volume bound. We modify Eq.~\ref{eq:K'} as follows:
\begin{equation}\label{eq:K'Mod}	
K'_{d}\equiv\ \sqcup_{\alpha}\alpha_{d}\, | \,\{ \,\omega_{\alpha_{d}}>\omega_0 \} .  \tag{\ref{eq:K'}\'}
\end{equation}
Analytical justification for this regulation is discussed in Sec.~\ref{sec:UV}.

\subsection{Cosmological Constant and Positive Definiteness}

The cosmological constant term also gives rise to divergences with $\Lambda\leq0$. Once we establish that $\omega_0>0$, it becomes immediate that $\Lambda<0$ would also lead to a network evolution into a state which is infinitely dense in disconnected points and an action which is unbounded by below. Any trivial point addition would be admitted, with $\Delta S=-|\omega_{0}\Lambda|$. For $\Lambda=0$, disconnected point addition would cause again $\Delta S=0$, as both curvature terms would be zero for the point volume contribution. We again argue as above that this is disallowed, and find that we are naturally restricted to $\Lambda>0$ as a combinatorial regime which supports compact network growth. Alternatively, we can view the cosmological and point volume terms as regulators, wherein only positive-definite values can admit actions which may be bounded from below and split the point-degenerate equivalence classes of action.

\subsection{Finite Probe Test for UV and IR Divergences}

As a simple toy model to probe for bounded changes in the action under Markov perturbations, we consider the addition of a single 1-simplex of weight $\omega_{1_1}$ with geometric attachment to an existing 1-simplex of weight $\omega_{2_1}$. The fluctuation here is given by 
\begin{equation}
\Delta S=\Lambda(\omega_{0}+\omega_{1_1})+2\omega_{0}\omega_{1_1}-\frac{\omega_{0}(\omega_{2_1}^{2}+\omega_{1_1}^{2})}{\sqrt{\omega_{2_1}\omega_{1_1}}}\,.
\end{equation}

In general, this expression has no definite sign. 

\subsubsection{UV \label{sec:UV}}

We can see that without the point volume regulator in the volumetric cutoff, a divergence to $-\infty$ would be present for an attachment of infinitesimal length, regardless of model parameters. With the cutoff, we forbid such a small length scale `ultraviolet' divergence and the change in action instead approaches the following: 
\begin{equation}
\lim_{\omega_{1_1}\rightarrow\omega_{0}}\Delta S=2\omega_{0}(\Lambda+\omega_{0}-\frac{(\omega_{2_1}^{2}+\omega_{0}^{2})}{2\sqrt{\omega_{2_1}\omega_{0}}})\,.
\end{equation}

In the mutual limit that both $\omega_{(1,2)_1}\rightarrow\omega_{0}$, we see that we recover the action equivalent of the addition of two isolated points, which matches our geometric intuition.

All cases of vanishing simplex addition then yield finite changes in the action, with the sign dependent on the model parameters and simplex lengths.

\subsubsection{IR}

The limit 
\begin{equation}
\lim_{\omega_{B}\rightarrow\infty}\Delta S\rightarrow-\infty
\end{equation} 
signals a large structure divergence in the `infrared' regime of the theory. Regardless of model parameters, a large enough simplicial probe attached to a 1-simplex will always yield a negative change in the action, and unbounds the global minimum of the theory. Understanding the nature of this divergence is the issue of current work in the model, as any system can allow a temporary fluctuation to generate the existence of an isolated 1-simplex, and by sending an infinite 1-simplicial probe, will immediately drop the system to a global minimum of maximally extended polymer-like geometries regardless of the initial state configurations.

\subsubsection{Bubble Divergences}

This IR divergence is extremely similar in nature to naive bubble and spike divergences found in spinfoam theories of quantum gravity. As a largely simplified explanation, a trivially satisfied constraint equation on the allowed irreducible representation labels in a closed subgraph of the foam admits an unbounded sum over all possible labelings of edges by half-integer representations  \cite{Baez2000}. This causes an explicit divergence of terms in the partition function. In such a case, a quantum-deformation of the gauge group can act to regulate the theory by restricting the number of good irreducible representations in the sum to a finite set and giving such a symmetry finite volume \cite{V.G.Turaev1992}. This deformation parameter is intrinsically tied to the existence of a positive definite cosmological constant.
\begin{figure}[H]
\centerline{\includegraphics[scale=2.6,angle=90]{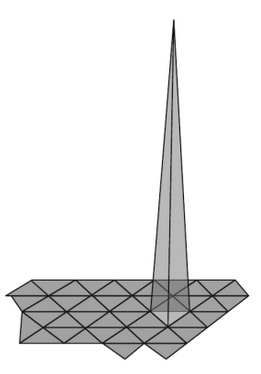}}
\caption{Cartoon Representation of a Spinfoam Spike Divergence \cite{Rovelli2015}}
\end{figure}

\subsubsection{Branching Universes}

Even with a naive IR divergence uncontrolled, such a behavior does not entirely invalidate the formalism. As the network evolution is accomplished by a series of finite Markov moves, we can simply restrict to a set of local moves which, on average, prevent such a divergence from materializing in practice. We examine instead for minima which, divergent moves unconsidered, act as meta-stable optimization points in topology, which can then be perturbed in volume to understand the local stability of the configuration. This is similar to the CDT handling of branching universes which would otherwise cause divergences in the state sum  \cite{Ambjorn2006}. The tendency for an infinite simiplicial probe may be interpreted as the emergence of a branching universe, where localized boundaries give rise to the birth of new geometry grown extremely non-locally. Seeing the same sort of divergence here may indicate that emergent geometry at the simplicial scale alone has such a property before we ever even consider a large geometry limit. 

The causes of such divergences in a CDT have roots in Regge calculus, where residual diffeomorphism symmetry in the bulk manifests as unconstrained translations of triangulation vertices \cite{Freidel2003}. Here, without contact to diffeomorphisms through even discretized manifolds, we still see such a divergence. Understanding our divergence as a possible manifestation of a kind of `combinatorial diffeomorphism' is underway. 

Nevertheless, unlike the typical characteristic of traditional Euclidean emergent geometry where the tension between the entropy of the state configuration plays against the unboundedness of the action to see a sharp phase change between either a crumpled or maximally extended state, we show that there exists tunable parameter regions where stable compact extended geometries can be still achieved as classical saddles \cite{Ambjorn2010}.
\begin{figure}[H]
\centerline{\includegraphics[scale=1.8]{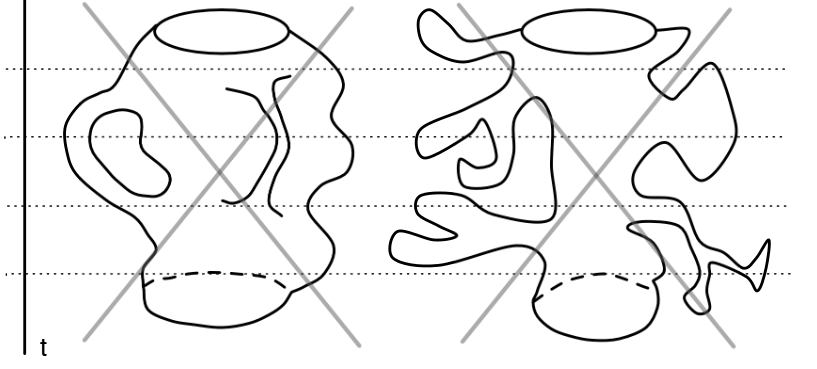}}
\caption{Cartoon Representation of Branching Universes in a CDT \cite{Loll2008}}
\end{figure}

\subsubsection{$\Lambda$ and Diffeomorphism Symmetry}

As mentioned, many of the naive infrared divergences in discrete emergent geometry models can be seen not as sicknesses, but as manifestations of residual diffeomorphisms on the vertexes of a fixed triangulation. This gauge volume can be rendered finite by the existence of a positive cosmological term. 

In a CDT, having a positive cosmological term corresponds to a maximum length scale through volume constraints. The topology is fixed to $S^3\times S^1$, and as each slice of the triangulation is constrained to be a triangulation of $S^3$, the bare cosmological constant is, on-shell, directly translated to the compact volume of the space as they are conjugate in the discretized action \cite{Ambjorn2010}. There are only two volumes of $4$-simplexes used in a $4$-D CDT, and as an average $4$-volume of a simplex can be defined and related to the $4$-volume of the Euclidean universe being simulated, the fixed topology gives a convenient bound on the maximum number of $4$-simplexes through the action. As such, CDT's are simulated at fixed space-time volume and fixed topology, and the sampling is constrained to sampling over triangulations with different reaches and volume \emph{distributions} (although an inverse Laplace transform can be used to relate the partition function to the variable volume case \cite{Ambjorn2005}). For a 3d spin foam formulation where the representation labels on the spin network edges correspond roughly to geometric edge lengths, the q-deformation cutoff can be reinterpreted geometrically proportionally to the finite maximum geodesic distance on a corresponding sphere of the same constant positive curvature \cite{Freidel2003}. Therefore, the gauge volume of residual diffeomorphisms as translations of spin network vertices giving an IR divergence is now finite. The symmetry is still present, but it at least has finite gauge volume which can be handled analytically. 

In the network gravity formalism, we introduced a cosmological constant as a combinatorial offset strictly as a regular on the number of disconnected components. We witness a naive IR divergence in our model that is not naturally regulated by mere existence of the cosmological term. It is not universally true for Euclidean geometries that simply because one includes a positive cosmological term, there exists a maximal length scale. Only on-shell can we relate $\Lambda$ and $R$ directly through the Einstein equations, and the local nature of these equations prevents us from making claims about any global structure like total volume. For example, one can introduce an $n$-cover of a sphere which will be locally indistinguishable from the base space with the same cosmological term and curvature, but has $n$-fold more volume.
Although there may be a way to interpret and regulate the divergence in the combinatorial model, without additional structure to understand the effect of a combinatorial cosmological term, such efforts may be independent of $\Lambda$. 

At present, our simulation heuristics demonstrate that the intrinsic scales of the boundary set the locality regulator for our optimization. If the simulation is allowed to probe scales much larger than the boundary scales, any structure in the boundaries is dwarfed by the globally unconstrained behavior of a free combinatorial-bulk. If the simulation is allowed to perturb within a compact region of the ambient space localized to the boundary scales, the simulation respects the boundary geometry and finds bounded minima. Understanding the nature of this relationship is at the heart of ongoing work.

\subsection{Refinement}

There is no sense in which we can explicitly take the number of simplexes to be very large or edge lengths to become small, given a fixed boundary. The algorithm naturally selects the number and size of simplexes as it iterates, and we have no control over the grown bulk once it has been established. Especially since we are probing structures that may not be related to triangulations in the far non-classical limit, we cannot use any assumptions applied from traditional discretizations. Therefore, we do not look to this model to necessarily generate a macroscopically smooth structure at any scale, but to understand the emergence of micro-geometry and its defects. There is the potential to understanding a refining limit holographically as imposed from the boundaries themselves, however.

One possibility that we intend to explore further is the existence of flat-action refinement. We detail some initial findings on flat-action dynamics in Sec.~\ref{sub:Action-Flat-Network-Dynamics}, giving analytical evidence for a state to be transformed within an equivalence class of action values and admitting the possibility to take a refining limit by translating laterally in action-space.

\begin{figure}[H]
\centerline{\includegraphics[scale=.3]{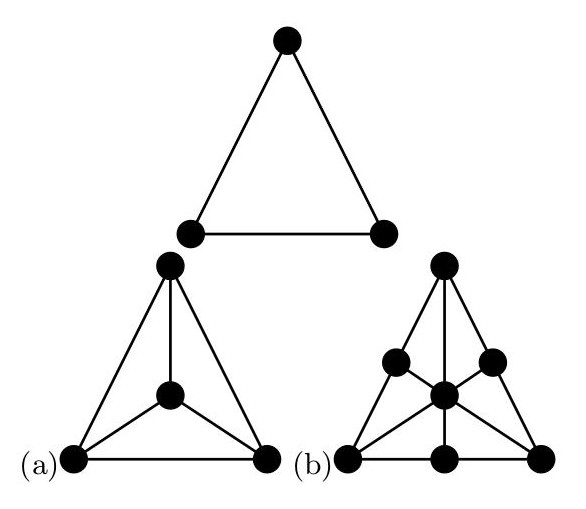}}
\caption{Refinement Example Under $1-3$ Pachner Move (a) vs Barycentric Subdivision (b)}\label{fig:BarycentricVsPachner}
\end{figure}

Lastly, a refinement procedure by conformal rescaling followed by barycentric subdivision (BSD) is being investigated. Unlike a triangulation where Pachner moves can be used to admit refinements or coarsening limits, due to our intersection pruning, such moves would not grant us a refining limit. For example, the $1-3$ Pachner move on a $2$-simplex would actually result in no $2$-simplexes under $\mu$ due to the outer triangle containing the inner triangles. A barycentric subdivision, however, provides the right nodal structure to be non-destructive under $\mu$ up to the minimal length scale, provided the initial state is geometric, as geometric realization is preserved under BSD \cite{Hatcher2002}. See Fig.~\ref{fig:BarycentricVsPachner} for an illustration of this procedure. 

\section{Observables}
Observables that can be considered are either purely topological or reference global properties. For example, $|K_{d}^{*}|$ or the number of path components $\pi_{0}$ can be calculated. An understanding of when a particular abstract complex is `approaching' a geometric complex can be gained from these observables. One method to measure this limit is to consider whether a particular complex can be reduced to a fully realized complex (even of uniform building blocks in the sense of a triangulation) in a finite number of additional pruning moves.

Of particular interest is the emergent dimensionality of the state, grown under various conditions (valence or embedding dimension restrictions), between various boundaries. One measure of dimension we aim to employ is the spectral dimension 
\begin{equation}
d_{s}=-2\lim_{\sigma\rightarrow\infty}\frac{\ln p_{\psi}(\sigma)}{\ln\sigma}\,,
\end{equation}
where $p_{\psi}(\sigma)$ is the return probability of a discrete diffusion of length $\sigma$.

Use of the spectral dimension as a diffeomorphism invariant measure related to physical dimension can be seen throughout the CDT literature \cite{Ambjorn2005}, since an unweighted diffusion is a purely combinatorial walk.

Taking the space of optimal embedded graphs as an ensemble itself, we can compute a different set of observables that can rely on local embedding data, such as the percentage of geometrically realized simplexes of a given order. These observables and the space of optimal graphs are interesting in their own right independent of the combinatorial space, as they have a more dense space of intersecting geometries to study and have an explicitly fully broken diffeomorphism symmetry, allowing us to probe observables such as Hausdorff dimensions averaged over ensemble embeddings.

\section{Simulation Results}

We restrict our discussion to the case when the coupling constants $\xi_{i}$ are all unity. 

\subsection{1-D Simulations and Cosmological Constant Driven Phase Transitions}
\begin{figure}[H]
\centerline{\includegraphics[scale=3]{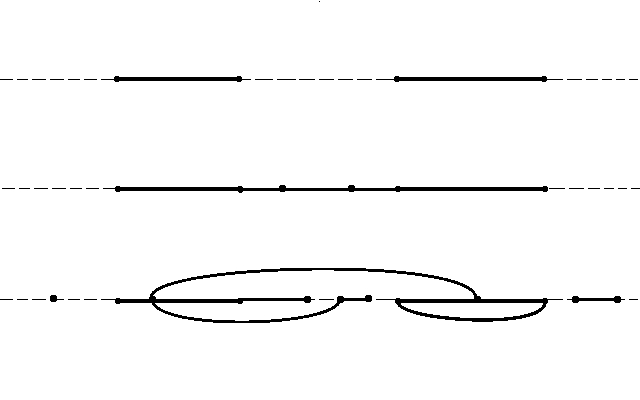}}\caption{From top to bottom, we illustrate the 1-D boundary setup; show an example geometric solution with simplicial attachments; and give an example abstract solution with non-simplicial attachments (shown with
curved lines for illustrating overlap) \label{fig:1dsetup}}
\end{figure}

A 1-dimensional embedding space is the simplest abstract network situation in which we can anticipate a classical state due to the restricted types of building blocks and attachments. Consider two closed disjoint intervals of the real line. Triangulate each interval with a single 1-simplex with 0-simplex boundaries. These networks can represent a $2$-boundary state that we would like to investigate.

A classical bulk network which connects these disjoint boundaries and represents a triangulation of an underlying manifold would be a series of 1-simplexes that connect the two segments in an embedded simplicial complex with non-overlapping simplicial volumes. However, in the space of embedded abstract simplicial complexes, we may admit isolated point additions (both overlapping edges or disjoint), and 1-simplex additions as disjoint, point-connected, overlapping, or doubly-point connected. The situation is illustrated in Fig.~\ref{fig:1dsetup}.

\begin{figure}[H]
\centerline{(a)\includegraphics[scale=3]{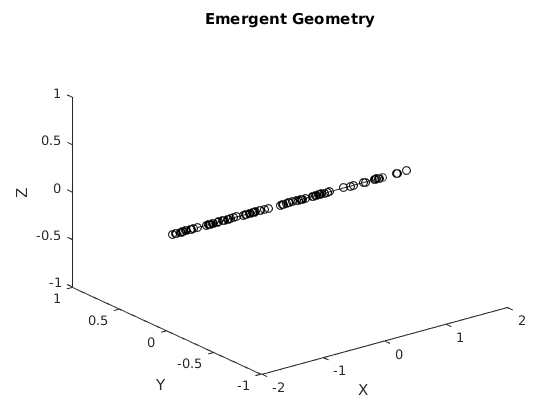}} \centerline{(b)\includegraphics[scale=3]{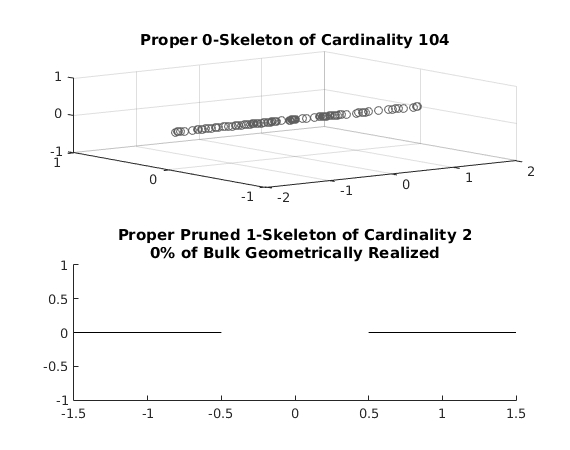}}
\caption{$\Lambda<0$ Optimal State (a) and $K^*$ Decomposition (b).
\protect \\
{The network growth fills all available space with nodes and does not produce fully geometrically realized images, with isolated nodes and overlaps in edges.} \label{fig:1DL-2}}
\end{figure}

\begin{figure}[H]
\centerline{(a)\includegraphics[scale=3]{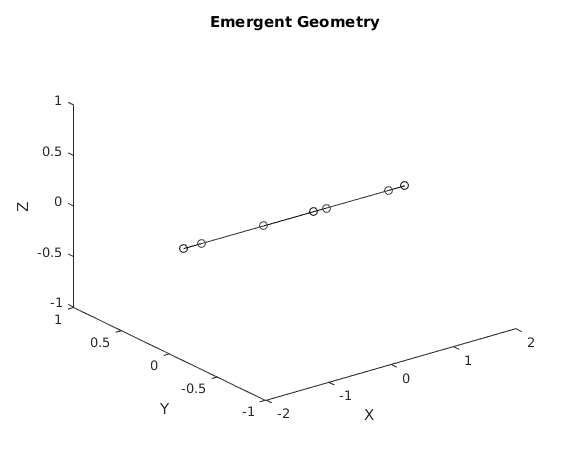}}\centerline{(b)\includegraphics[scale=3]{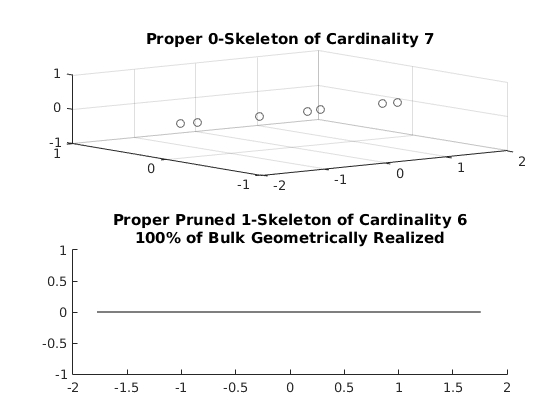}}
\caption{$0<\Lambda<\Lambda_{c}$ Optimal State (a) and $K^*$ Decomposition (b). \protect \\
{We see the minimum in the action appears with the emergence of simplicial attachments that have proper geometric embeddings admitting a perfect skeletal decomposition that lacks any unrealized or disconnected geometries.} \label{fig:1DL2}}
\end{figure}

\begin{figure}[H]
\centerline{(a)\includegraphics[scale=3]{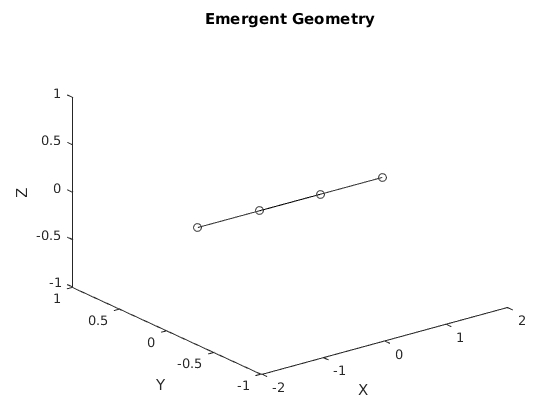}}\centerline{(b)\includegraphics[scale=3]{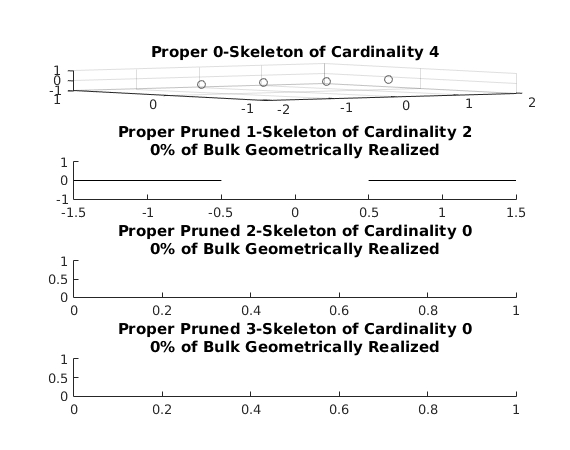}}
\caption{$\Lambda>\Lambda_{c}$ Optimal State (a) and $K^*$ Decomposition (b). \protect \\
{We see that no geometric network emerges--the action began in its global minimum with the boundary configuration, and lateral translations into degenerate configurations induce an over-connected graph state.} \label{fig:1DL4}}
\end{figure}

Example optimal states for uniform coupling and positive point volume are shown in Figs.~\ref{fig:1DL-2}, \ref{fig:1DL2}, and \ref{fig:1DL4}. We see that for negative cosmological constant the system begins to fill all available space with nodes. This echoes our earlier analysis of evolution in the presence of a non-positive-definite cosmological constant. For a positive definite cosmological term with a value less than a particular critical value $\Lambda_{c}$, the simulation demonstrates a geometric complex solution we would anticipate as a minima of the action, with fully realized skeletons. Lastly, when probing the system for $\Lambda>\Lambda_{c}$, we see that no realizable network is favored to grow as the system began in its global minimum. The system is perturbed into a highly connected state, but as keeping or removing intersecting geometries are equivalent under this parameter set, we are left with an over-connected graph state on the original set of nodes. The critical value $\Lambda_{c}$ was found through numerical investigation, and does not yet have an analytical justification; nevertheless, $\Lambda$ acts as a clear parameter for controlling the phases between very different states of stochastic network growth, focused on either the network connectivity or abundance of structural elements. This indicates the importance of the cosmological term as a regulator of non-degenerate state growth with respect to fixed boundaries.


\subsection{Higher Dimensional Simulations}
\begin{figure}[H]
\centerline{(a)\includegraphics[scale=3]{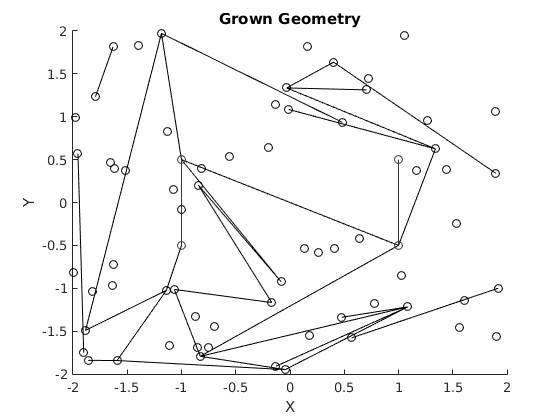}}\centerline{(b)\includegraphics[scale=3]{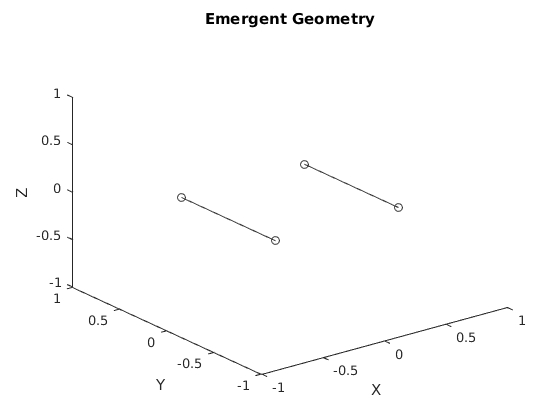}}
\centerline{(c)\includegraphics[scale=3]{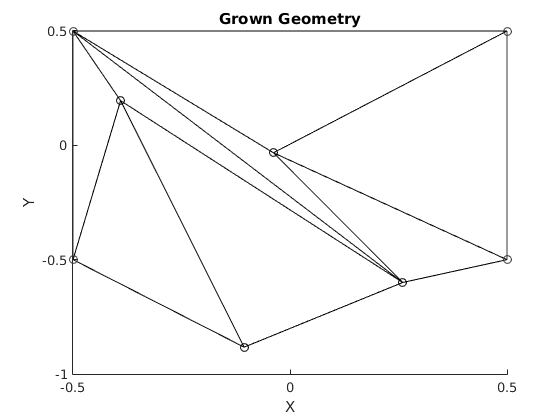}}
\caption{Geometry Restricted to $\mathbb{R}^{2}$ with Boundary States of Two Finite 1-Simplexes.\protect \\
{We see the characteristic phases of negative (a) or super-critical (b) cosmological growth, with either chaotic elements with disconnected regions or no network growth at all, respectively; in c), we see realized growth for the geometric cosmological range, forming a completely realized $2$-triangulation.}\label{fig:2DSim}}
\end{figure}
\begin{figure}[H]
\centerline{(a)\includegraphics[scale=3]{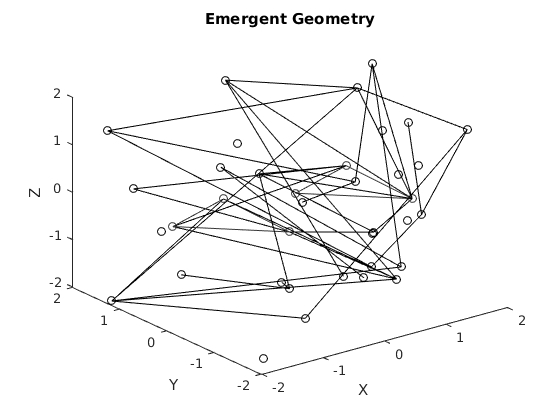}}\centerline{(b)\includegraphics[scale=3]{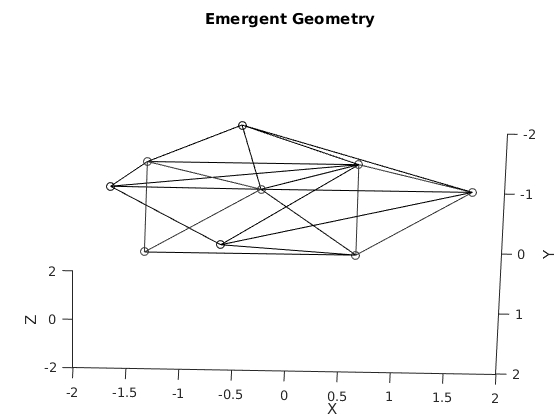}}
\caption{Identical $2$-Simplex Boundaries and their Network Growth for Negative (a) and Geometric Cosmological Parameters (b) \protect \\
{We can clearly see that the disconnected and disordered network of the negative cosmological simulation starkly contrasts with the positive-subcritical simulation.}\label{fig:3DSim}}
\end{figure}
We provide examples of a higher dimensional embedding simulations in Fig.~\ref{fig:2DSim} and \ref{fig:3DSim}, illustrating final network growths for $\Lambda$ in the identified phases.

Although the presented simulation results are anecdotal, these examples do demonstrate that the action and growth procedure are capable of investigating the emergence of realized geometries without preassigning attachment rules or dimensional assumptions. In the generic space of abstract simplicial complexes, it is a highly nontrivial goal to find a phase of stochastic growth with geometric realization and compact yet voluminous extent. Although there is certainly further work to be done, that the model can naturally demonstrate precisely such a phase, and even generate genuine triangulations as classical minima, is strongly encouraging as a base framework. 
Systematic investigation into geometric phases is underway.

\section{Action-Flat Network Dynamics\label{sub:Action-Flat-Network-Dynamics}}

We restrict ourselves now to the case of model interest, where $(\omega_0,\Lambda)>0$. 

In this regime, we ask whether there are equivalence classes under the action for various states. To begin our analysis and simplify the investigation, we restrict to first probe independent regular simplicial building blocks. This provides us with a 3-dimensional configuration space of variables: $\omega_0$, $\Lambda$, and $\omega$ denoting the uniform edge length.

We can now easily generalize to the case of an arbitrary regular $n$-simplex of side-length $\omega$, and find the action in closed form: 
\begin{eqnarray}
\omega_{d} & = & \frac{\omega^{d}}{d!}\sqrt{\frac{d+1}{2^{d}}}(1+\delta_{0}^{d}(\omega_{0}-1))\\ \nonumber
S_{n} & = & \sum_{d=0}^{n}\binom{n+1}{d+1}\omega_{d}\left\{ \Lambda+(1-\delta_{0}^{d})\right.\\
 & \times & \omega_{d}\left((n-d)\frac{\omega_{d}}{\omega_{d+1}}+(d+1)\frac{\omega_{d-1}}{\omega_{d}}\right.\nonumber \\
 & - & \left.\left.(2n-1-d)\Theta(2n-1-d-n)|\frac{\omega_{d}}{\omega_{d+1}}-\frac{\omega_{d-1}}{\omega_{d}}|\right)\right\} \nonumber 
\end{eqnarray}
Here, the $\delta_{\cdot}^{\cdot}$ is the Kronecker delta, and $\Theta(\cdot)$ is the Heaviside function with the imposition that $\Theta(0)=0$.

\begin{figure}[H]
\begin{centering}
\includegraphics[scale=4]{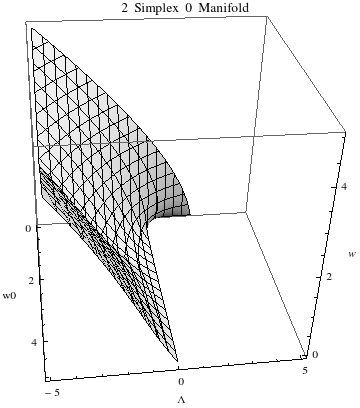}
\par\end{centering}
\caption{$S=0$ manifold as a function of $(\omega,\omega_{0},\Lambda)$ for an isolated 2-simplex, with the first indications of multi-valued behavior for fixed model parameters as we vary $\omega$ }\label{fig:reg2Simp0}
\end{figure}

We study the level-sets of this function, solving for the roots of this action as an example. We see that for progressively higher $n$ simplexes, the solution space has larger forbidden regions in the parameter space, and even discontinuous regions, but nevertheless the space of solutions does not appear to become discrete or vanish. 
\begin{description}
\item [{2-Simplex}] The regular 1-simplex action has a single isolated point as level sets and is uninteresting. We look toward the 2-simplex action, which takes the more complicated form: 
\begin{eqnarray}
S&=&3\omega_{0}\Lambda+\frac{\sqrt{3}}{4}\omega^{2}(3\omega+\Lambda)\nonumber \\ &+& \omega(4\sqrt{3}+6\omega_{0}+3\Lambda-2\omega|\frac{4\sqrt{3}-3\omega_{0}}{\omega}|)\,.
\end{eqnarray}
We can see in Fig.~\ref{fig:reg2Simp0} that there does exist a solution set for $S=0$ and $\Lambda>0$ (this plot extends to both $\pm\Lambda$ in the interest in seeing the structure). Thus, certain values of cosmological and point volume terms admit the spontaneous emergence of 2-simplexes for even a purely classical action descent. Furthermore, we see that $\omega$ appears cubically in the above equation, allowing for the possibility of a multivalued solution. This indicates that a spontaneous transition between regular 2-simplexes of varying size is admitted by the action as well, for particular regions of the free parameter space.
\end{description}

\begin{figure}[H]
\begin{centering}
\includegraphics[scale=4]{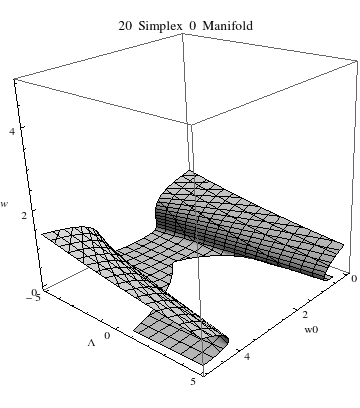}
\par\end{centering}
\caption{$S=0$ manifold as a function of $(\omega,\omega_{0},\Lambda)$ for an isolated 20-simplex, clearly showing a rich parameter space of multivalued solutions in $\omega$ }\label{fig: reg20Simp0}
\end{figure}

\begin{description}
\item [{n-Simplex}] We have continued to probe higher regular simplex structures for their flat manifolds with respect to the configuration space, and can report a similar behavior. We provide an example in Fig.~\ref{fig: reg20Simp0} of a regular 20-simplex $S=0$ manifold for evidence of multivalued behavior for some parameter regimes, as well as large regions where there is clearly no such degenerate behavior. 
\end{description}

The existence of these manifolds suggests equivalence classes of networks under the action, where the network dynamics may translate laterally along these flat directions before jumping into a new region of the state space. The states themselves are distinct elements in the state space. However, for the purposes of numerical simulation, the possibility for a network to deform along a flat parameter manifold before jumping into a state with a radically different action compared to the primary state introduces a new level of complexity. The system can be prone to large fluctuations in the action, and sampling efficiently can be difficult with non-isolated degeneracies and potential domain walls.

It should be noted that, for a given simulation, the parameters of $\omega_0$ and $\Lambda$ are non-dynamical. We also clearly do not expect only regular isolated simplexes to constitute the states. Nevertheless, it is still unknown whether the properties seen in this restricted numerical investigation will be present in the general state case, or whether the flat manifold degeneracies are split into a sufficiently discrete space. 

The proof of existence and study of behavior of action-degenerate general networks is underway. 

\subsection{Trivial Simplex Transitions }

Echoing the above investigation into regular simplex action-flat manifolds, we also examined whether there exists \emph{different} order simplexes which yield the same action under edge length variation, corresponding to transitions between distinct topologies which costs the system zero energy. The answer to the existence of these transitions is in the affirmative--depending on the parameter space, these transitions may or may not be forbidden, and there may even exist multiple degenerate configurations. 

\begin{figure}[ht]
\centerline{\includegraphics[scale=4]{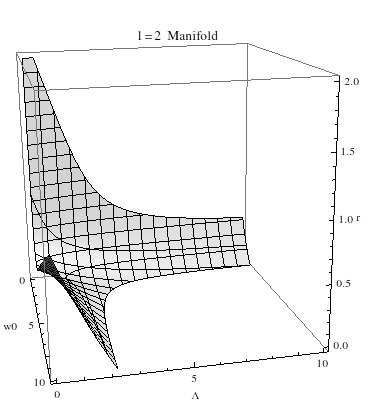}}
\caption{Action-equivalent manifolds for isolated 1-2-simplexes as a function of $(r,\omega_{0},\Lambda)$. In regions of this parameter manifold, 1-simplexes may spontaneously transition into 2-simplexes of different sizes for fixed model parameters. }\label{fig:1-2-Equivalence}
\end{figure}

\begin{figure}[ht]
\centerline{\includegraphics[scale=4]{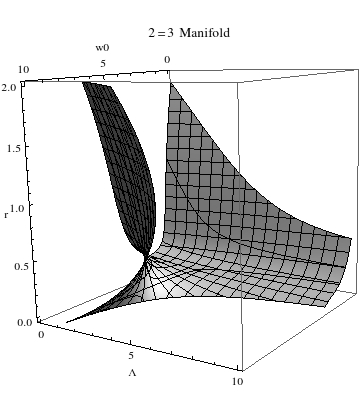}}\caption{Action-equivalent manifolds for isolated 2-3-simplexes as a function of $(r,\omega_{0},\Lambda)$. In regions of this parameter manifold, 2-simplexes may spontaneously transition into 3-simplexes of different sizes for fixed model parameters. }\label{fig:2-3-Equivalence}
\end{figure}

Setting the base simplex to have unit length edges and probing the configuration space $(\omega_{0},\Lambda,r)$, where $r$ is the scaling dimension for the compared simplex, we can search for solutions where $S(n\text{-simplex})=S(m\text{-simplex})$ . Figs.~\ref{fig:1-2-Equivalence} and \ref{fig:2-3-Equivalence} show the critical surfaces for such solutions.


\section{Discussion}

The study of emergent complex networks is a largely active field in which a variety of network growth paradigms are investigated for behavior which can be mapped onto physical problems of interest \emph{a posteriori}. Rules which govern network growth in a given model are often selected without a motivating principle, as it is difficult to gain a thorough understanding of the emergent properties of the model without fully running the stochastic growth. 

We have presented an emergent network formalism with strong analytic handles which can be used to investigate the nature of network interactions and probe emergent geometry considerations as mapped onto a combinatorial state space of interest. In building a gravitationally motivated model to study outstanding questions in simplicial emergent geometry, we allowed for a stochastic Markov process to generate a bulk network state between fixed boundary configurations, sampling for the stable minima of the semi-classical state space through an annealed Metropolis algorithm in order to probe our combinatorial measure. The metric against which we evaluated the network growth is an action principle in the space of abstract simplicial complexes, utilizing a decomposition of the embedded state into a superposition of proper skeletons consisting of geometrically realizable simplexes with non-empty simplicial volumes. Using the Forman cellular curvature with combinatorial weights provided by simplicial volumes, we showed that a Regge-like network action demonstrates distinct phases of stochastic geometric growth driven by the cosmological constant without preassigning the dimension of the simplicial building blocks, attachment rules, or dedicated embedding dimension. With fewer constraints imposed by hand, we connected the qualitative behavior of the model to existing frameworks for emergent geometry and aspects of discrete quantum gravity, and illustrated numerical and analytic justifications for a positive-definite cosmological constant and minimum combinatorial length scale.

Upcoming work aims to answer some of the many questions this construction poses, and begin to systematically compute observables for a wide variety of configurations in addition to demonstrating analytical proofs for the tractable sectors of the combinatorial space.

We would lastly like to mention that in addition to studying geometric embeddings from combinatorial structures, this formalism may be able to be used to model a variety of other types of network dynamics. For example, a study of optimal social networks, neural networks, or networks related to commerce can be mapped onto this formalism. We can furthermore consider extensions such as adding external potentials in the ambient space to guide embedded network growth, or introduce combinatorially static sub-networks which can be passed freely within the bulk without a fixed anchor as `matter' analogues.

There is a rich landscape to study, and the framework presents a numerical `petri-dish' of emergent combinatorial geometry to probe. 

\begin{acknowledgments}

This work is supported in part by the University of Washington. Special thanks are given to the GitHub Education branch, for providing a student developer package and hosting of the Network Gravity repository; to Lee Smolin and Laurent Freidel, for inviting me to the Perimeter Institute and our conversations on discrete quantum gravity and the introduction to networks which prompted this investigation; and to Stephen Sharpe, for enabling my initial excursion into the wild world of quantum gravity and continuing to be a tremendous source of encouragement and optimism.

This research was supported in part by Perimeter Institute for Theoretical Physics. Research at Perimeter Institute is supported by the Government of Canada through Industry Canada and by the Province of Ontario through the Ministry of Economic Development \& Innovation.
\end{acknowledgments}

\appendix

\section{Gauge Symmetries\label{sec:Symm}}

In terms of inherited symmetries under $\mu$ from the embedded state, foremost the action is invariant under $ISO(m)$. Additionally there is the group of symmetries which maintains the combinatorial weights and adjacency data but deforms the underlying coordinate space. This can be seen as a space of restricted diffeomorphism, where one can freely deform points (actively or passively) as long as edge lengths as measured from the induced metric on the ambient space are invariant between all \emph{connected} nodes. An example would be the case of two 1-simplexes attached at a common node. Without changing the action, either end-node can be displaced by pivoting around the central node on some $S^{m}$ with a radius of the embedded edge length (modulo degeneracy considerations). The same can be said for any structure with a higher dimensional `pivot'. The number of free pivots characterize the continuous symmetry in the available ambient space and creates a large class of equivalent states under the action up to such coordinate displacements.

We also have symmetries of the graph state pre-embedding that are shared by the combinatorial state space alone. Writing down the state in an adjacency matrix sets an initial labeling, but we have vertex automorphisms $AUT(g)$ which preserve the adjacency structure but permute the labels and give the same combinatorial data. More importantly, the combinatorial state is truly labeling independent. The number of weighted-cliques of a given order and how many neighbors they each have make up the data alone. A complete relabeling of the graph state can give the same data, and this symmetry is larger than just the vertex automorphisms.

\section{Combinatorial Inequalities\label{sec:ComboIneq}}

For a fixed $K^{*}$ structure, there are determinant inequalities on the weight structures from Eq.~\ref{eq:omega} to insure that there exists some simplicial volume greater than $\omega_{\alpha_0}$ provided a range of edge lengths, and inequalities that come from embeddability. The embeddability requirements are the most difficult to characterize. These dictate, for example, that for any closed $m-1$-surface $A$ in an ambient $m$-space, every permutation of sums of the surface-weights with a single element removed must be greater than the weight of the removed element:
\begin{equation}
 (\sum_{\alpha_{m-1}\neq \hat{\alpha}_{m-1}\in K^{*}_{m-1}|_A} \omega_{\alpha_{m-1}}) > \omega_{\hat{\alpha}_{m-1}}\,\forall \,\hat{\alpha} \,.
\end{equation}

These include the higher dimensional closure inequalities on the weights for given simplexes by taking the surface to be a single element (like the triangle inequalities), but also extend to surfaces which live entirely in sub-dimensional 
hyperplanes. There are further constraints which are even harder to characterize: for example, given a vertex in $2$-D that is surrounded by simplically attached $2$-simplexes except for a small deficit, one needs inequalities on the allowed $2$-simplex which would only share the internal vertex and live in the remaining ambient `wedge', which is a highly non-local and non-trivial inequality. 

Furthermore, the existence of combinatorial boundaries presents equality constraints on the allowed level-$1$ weights connecting boundary roots, which can be seen more clearly in the embedded picture as constraints due to the relative positioning of the fixed boundary structures. 

A full understanding of when an abstract simplicial complex can be embedded in a fixed dimensional Euclidean ambient space with prescribed simplicial volumes is tractable for $1$ dimension and possibly for $2$ dimensions, but immediately becomes unwieldy for higher dimensional spaces.

\section{Knill Curvature \label{sec:Knill-Curvature}}

An oft used curvature in network theory is the Knill curvature, defined for simple undirected networks $g$ at a given node $v$ by counting the number of complete subgraphs of order $i$ attached to said node (here denoted $\#_{v}^{i}$)  \cite{Wu2015,Knill2011}.

\begin{subequations}
\begin{eqnarray}
R_{v}&=&\sum_{i=1}^{\infty}\frac{(-1)^{i+1}}{i}\#_{v}^{i}\,, \\
\chi(g)&=&\sum_{v\in V}R_{v}\,.
\end{eqnarray}
\end{subequations}

We see that when summed over the network, the Euler character is returned, which gives a discrete analog of the Gauss-Bonnet theorem for networks. We avoid using the Knill curvature for two reasons. Foremost, this curvature form is purely combinatorial and does not take into account relative weighting between complete graphs of different sizes. Although trivially diffeomorphism invariant, this curvature is at odds with the notion of curvature we would like to associate with our model where intrinsic geometric data is accounted for. Secondly, while the Knill curvature as an Euler density appears similar to the Ricci curvature for 2 dimensions, this analogy clearly breaks down in higher dimensional Riemannian manifolds where the Ricci curvature is no longer a topological density. The Knill curvature is always measuring topological properties, and we seek a measure of curvature which is not strictly always an Euler density.

\bibliography{MyLibrary}

\end{document}